\documentclass[amsmath,amssymb, 12pt]{revtex4}

\usepackage{graphicx}
\usepackage{epsfig}
\usepackage{bm}

\makeatletter \leftline{\epsfbox{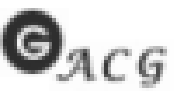}}
 \leftline
{\footnotesize{\textbf{\textsf{GACG-04/10}}}}

\begin{document}
\newcommand{\gmn}{ g_{\mu \nu} }
\newcommand{\beq}{\begin{equation}}
\newcommand{\eeq}{\end{equation}}
\newcommand{\sst}{\scriptscriptstyle}
\newcommand{\ds}{\displaystyle}
\newcommand{\ddd}{\Phi_{,{\rho}}\Phi^{,{\rho}}}
\newcommand{\re}{{\frac{1}{2}g_{\mu\nu}}R}
\newcommand{\trr}{\Phi_{,{\mu}}\Phi_{,{\nu}}}
\newcommand{\be}{\begin{equation}}
\newcommand{\en}{\end{equation}}
\newcommand{\bea}{\begin{eqnarray}}
\newcommand{\ena}{\end{eqnarray}}
\newcommand{\des}{\baselineskip 24pt}
\topmargin -0.5cm

\title{Curvature in causal BD-type inflationary cosmology.}
\author{Francisco Pe\~na$^1 \thanks{E-Mail: fpena@gacg.cl}$,
Mauricio Cataldo$^2$\thanks{E-Mail: mcataldo@ubiobio.cl} and
Sergio del Campo$^3 \thanks{E-Mail: sdelcamp@ucv.cl}$
  }
  \address{$^1$ Departamento de Ciencias F\'\i sicas, Facultad de
Ingenier\'\i a, Ciencias y Administraci\'on, Universidad de la
Frontera, Avda. Francisco Salazar 01145, Casilla 54-D, Temuco,
Chile.}
\address{$^2$ Depto. de F\'\i sica, Facultad de Ciencias, Universidad del B\'{\i}o B\'{\i}o, \\
Av. Collao 1202, Casilla 5-C, Concepci\'on, Chile.}
\address{$^3$ Instituto de F\'\i sica, Universidad Cat\'olica de
Valpara\'\i so, \\ Av Brasil 2950, Valpara\'\i so, Chile.\\}


\begin{abstract}
We study a closed model of the universe filled with viscous fluid
and quintessence matter components in a Brans-Dicke type
cosmological model. The dynamical equations imply that the
universe may look like an accelerated flat
Friedmann-Robertson-Walker universe at low redshift. We consider
here dissipative processes which follow a causal thermodynamics.
The theory is applied to viscous fluid inflation, where accepted
values for the total entropy in the observable universe is
obtained.\vspace{0.5cm}

\noindent {\it {}}
\end{abstract}


\maketitle

\section{\bf Introduction}
Recent observation of the Hubble diagram for supernovae Ia
indicates that the expansion of the universe is acceleranting at
the present epoch \cite{Perlmutter,Riess}. This acceleration is
attributed to a dark energy residing in space itself, which also
balances the kinetic energy of expansion so as to give the
universe zero spatial curvature, as deduced from the cosmic
microwave background radiation\cite{Siviers}.

Initial interpretation was to consider that in the universe there
exists an important matter component that, in its most simple
description, has the characteristic of the cosmological constant
$\Lambda$, i.e. a vacuum energy density which contributes to a
large component of negative pressure, and thus accelerates rather
than decelerates the expansion of the universe. An alternative
interpretation is to consider quintessence dark energy, in the
form of a scalar field with self interacting
potential\cite{CaDaSt}.

On the other hand, in recent years an important attention has
received cosmological models in which bulk viscosity is
considered. For example, it was shown that the introduction of
this kind of viscosity into cosmological models can avoid the big
bang singularities\cite{Murphy,Belinkskii,Golda} and any
contribution from particle production may be modelled as an
effective bulk viscosity\cite{Maartens}. The bulk viscosity arises
typically in mixtures of different species (as in a radiative
fluid) or of the same (but with different energies) fluids. The
dissipation due to bulk viscosity converts kinetic energy of the
particles into heat, and thus one expects it to reduce the
effective pressure in an expanding fluid. This fact may play a
crucial role in the inflationary era of the universe since it is
interesting to know whether dissipative effects could be strong
enough to do a large negative effective pressure leading to
inflation.

Many, probably most, of the inflationary cosmological models
considered are of flat Friedmann-Robertson-Walker (FRW) type. In
this kind of models the velocity gradients causing shear viscosity
and temperature gradients leading to heat transport are absent.
Thus, any dissipation in FRW universe may be modelled as a bulk
viscosity within a thermodynamical approach. These  dissipative
processes, that consider the bulk viscosity\cite{4,5}, are
compatible with the homogeneity and isotropy assumptions for the
universe, and may play an important role in the early universe,
specially before nucleosynthesis\cite{Gron1}.

Curvature models have been studied with an important matter
component whose equation of state is given by $p = - \rho / 3$.
Here, the universe expands at a constant speed\cite{Ko}. On the
other hand, a flat decelerating universe model has been
described\cite{CrdCHe}. Also, following a similar idea in
accelerating models, the dark energy potential\cite{SdC1} and the
location of the first Doppler peak in the Cosmic Microwave
Background spectrum\cite{SdC2} have been studied. In this way, we
could consider models where the starting geometry were other than
that corresponding to the critical geometry, but at low redshift
these are indistinguishable from flat geometries.

In this paper we want to study universe models in which we use the
causal thermodynamical theory for processes out of
equilibrium\cite{Israel}. The stable and causal thermodynamics of
Israel and Stewart replaces satisfactorily the unstable and
non-causal theory of Eckart\cite{Eckart} and Landau and
Lifshitz\cite{Landau}. The main goal in this paper is to study not
vanishing curvature causal inflationary universe models. We
pretend to do this in Brans-Dicke (BD)-type theory\cite{BD}. Here
the theory is characterized by a scalar field $\Phi$ and a
constant coupling function $\omega_o$ together with a scalar
potential associated to $\Phi$. From now on the subscript zero
will represent the actual values.

\section{\bf {The field equations}}
We consider the effective action to be
\begin{eqnarray}
\label{accion} S=\int{d^4 x}\sqrt{-g}\{\frac{1}{16\pi}(\Phi R -
\frac{\omega_0}{\Phi}(\partial_{\mu} \Phi)^{2} + V(\Phi)) +
\frac{1}{2}(\partial_{\mu} Q)^{2} + V(Q) + {\cal{L}}_{\sst M}\}
\end{eqnarray}
where $\Phi$ is the BD scalar field,$V(\Phi )$ is a scalar
potential associated to the BD field, $\omega_{0}$, is the BD
parameter, $Q$ the quintessence scalar field with associated
potential $V(Q)$, $R$ the scalar curvature and ${\cal{L}}_{M}$
represents the matter contributions other than the $Q$ component.

We will obtain the field equations for a cosmological viscous
model where the energy-momentum tensor of the fluid with a bulk
viscosity is taken to be
\begin{eqnarray}
\label{Tmateria} T_{\mu \nu }=(\rho_{\sst{M}}+ p_{\sst{M}}+\Pi
)v_{\mu }v_{\nu }+(p_{\sst{M}}+ \Pi )g_{\mu \nu },
\end{eqnarray}
with $\rho_{\sst{M}}$,  $p_{\sst{M}}$  the thermodynamical
density, the pressure of the fluid, and $\Pi$ is the bulk viscous
pressure, respectively. In the co-movil reference system we take
$v^{\mu}v_{\mu}=-1$ and $v^{\mu}=\delta_{0}^{\mu}$.

The contribution to the energy-momentum tensor related to the
quintessence scalar field, becomes given by:
\begin{eqnarray} \label{Tquinta} T_{\mu \nu}=(\rho_{\sst{Q}}+
 p_{\sst{Q}})v_{\mu }v_{\nu }+g_{\mu \nu }p_{\sst{Q}},
\end{eqnarray}
We can get the field equations of the universe from the
action(\ref{accion}), to the FRW metric.
\begin{eqnarray}
\label{00FRW} H^2+H\frac{\dot{\Phi}}{\Phi}=\frac{\omega_0}{6}
\left(\frac{\dot{\Phi}}{\Phi}\right)^2+\frac{8\pi}{3\Phi}
(\rho_{\sst{M}}+\rho_{\sst{Q}})-\frac{k}{a^2}+\frac{1}{6\Phi}V(\Phi)
\end{eqnarray}
$$
 2\dot{H}+3H^2+\frac{k}{a^2}=-\frac{\omega_{0}}{2}
\left(\frac{\dot{\Phi}}{\Phi}\right)^2-\frac{1}{\Phi}(\ddot{\Phi}+2H\dot{\Phi})+
$$

\begin{eqnarray}
\hspace{3.0cm}\label{ijFRW}\frac{V(\Phi)}{2\Phi}-\frac{8\pi}{\Phi}(p_{\sst{
M}}+p_{\sst{Q}}+\Pi)
\end{eqnarray}

$$
\hspace{-1.5cm}\label{BDeq}\ddot{\Phi}+3H\dot{\Phi}+\frac{\Phi^3}{2\omega_{0}+3}\frac{d}{d\Phi}\left(\frac{V(\Phi)}{\Phi^2}\right)=
$$
\begin{eqnarray}
\hspace{2.0cm}\frac{8\pi}{2\omega_{0}+3}(\rho_{\sst{M}}-3p_{\sst{M}}+\rho_{\sst{Q}}-3P_{\sst{Q}}-3\Pi)
\end{eqnarray}
\begin{eqnarray}
\label{Qeq}\ddot{Q}+3H\dot{Q}=-\frac{\partial V(Q)}{\partial Q}
\end{eqnarray}
where $a(t)$ is the scale factor, and $k$ is the curvature
parameter that is equal to 1 (closed), 0 (flat) or -1 (open).

 The matter conservation equation is given by
\begin{eqnarray}
\label{rhofluido} \dot{\rho}_{\sst M}+3H(\rho_{\sst M}+p_{\sst
M}+\Pi)=0
\end{eqnarray}
where the matter has a barotropic equation of state
$p_{\sst{M}}=(\gamma -1)\rho_{\sst{M}}$ where the parameter
$\gamma$ is in the ranger $0\leq\gamma\leq3$. The energy density
associated to the quintessence scalar field is given by,
\begin{eqnarray}
\label{rhoquinta} \rho_{\sst{Q}}=\frac{1}{2}{\dot{Q}}^{2}+V(Q)
\end{eqnarray}
and the pressure
\begin{eqnarray}
\label{pquinta} p_{\sst{Q}}=\frac{1}{2}{\dot{Q}}^{2}-V(Q).
\end{eqnarray}

This quintessence scalar field has a equation of state defined by
$p_{\sst{Q}}=w_{\sst{Q}}\rho_{\sst{Q}}$, where $w_{\sst{Q}}$ is
the eq. of it parameter, and similar to the matter fluid, it
satisfies the conservation equation
\begin{eqnarray}
\label{conservacion del campo escalar}
\dot{\rho}_{\sst{Q}}+3H(\rho_{\sst{Q}}+p_{\sst{Q}})=0.
\end{eqnarray}

\section{Characteristics of the model}

In order to mimic a flat universe model, we impose the following
conditions\cite{CrdCHe}.
\begin{eqnarray}
\label{plano1}\frac{8\pi}{3\Phi}\rho_{\sst{Q}}-\frac{k}{a^2}+\frac{1}{6\Phi}V(\Phi)=0
\ena and
 \bea \label{plano2} \rho_{\sst{Q}}=\frac{\Phi^3}{8\pi
(1-3\omega_{\sst{Q}})}\frac{d}{d\Phi}\left(\frac{V(\Phi)}{\Phi^2}\right).
\end{eqnarray}
In the following we will assume a power law potential for the BD
field, $V(\Phi)=V_{0}(\frac{\Phi}{\Phi_{0}})^\beta$, where $V_0$
and $\beta$ are constant. We will see that $\beta$ becomes
determined entirely by the equation state parameter
$\omega_{\sst{Q}}$.  With this assumption we obtain from the two
latter eqs.
\begin{eqnarray}
\label{plano3}\frac{k}{a^2}=\left(\frac{2\beta -
3(1+w_{\sst{Q}})}{6(1-3w_{\sst{Q}})}\right)
\frac{V_{0}}{\Phi_{0}}\left(\frac{\Phi}{\Phi_{0}}\right)^{\beta -
1},
\end{eqnarray}
where for $V_{0} > 0$ and $\Phi_{0} >0$, we obtain that $\beta >
\frac{3}{2}(1+w_{\sst{Q}})$ if $k>0$,
$\beta=\frac{3}{2}(1+w_{\sst{Q}})$ if $k=0$ and $\beta <
\frac{3}{2}(1+w_{\sst{Q}})$ if $k<0$. Since we are interested in
power law inflationary universe models, then we take the scale
factor to be given by
\begin{eqnarray}
\label{factor de escala} a(t)=a_0\left(\frac{t}{t_0}\right)^{N}.
\end{eqnarray}
with $N>$1. Therefore, we obtain for the BD field
\begin{eqnarray} \label{BD sol} \Phi(t)=\Phi_{0}\left(\frac{t}{t_0}\right)^{\frac{2N}{1-\beta}}
\end{eqnarray}
and
$V_0=\frac{6k(1-3w_{\sst{Q}})}{2\beta-3(1+w_{\sst{Q}})}\frac{\Phi_0}{a_0^2}$.
With these solutions, the quintessence energy density results in
\begin{eqnarray}
\label{rhoQ
sol}\rho_Q=\frac{6k(\beta-2)}{8\pi(2\beta-3(1+w_{\sst{Q}}))}
\frac{\Phi_0}{{a_0}^2}\left(\frac{t}{t_0}\right)^{\frac{2N\beta}{1-\beta}}
\end{eqnarray}
Since $N>1$, then we need $\beta<0$ or $\beta >1$  in order to
satisfy the energy conservation equation. From the $Q$ field
eq.(\ref{conservacion del campo escalar}) this gives
$w_{\sst{Q}}=-\frac{1}{3}\frac{3-\beta}{1-\beta}$, or equivalently
$\ds \beta = 3 \frac{1+w_{\sst{Q}}}{1+ 3w_{\sst{Q}}}$, which
becomes determined in terms of the $w_{\sst{Q}}$ parameter as was
mentioned above. We should note that since $-1<w_{\sst{Q}}<-1/3$,
then $\beta <0$, therefore the possibility of satisfying $\beta>1
$ is not possible. This implies that the BD scalar potential is an
inverse power law potential of the BD field.

The field solutions together with the constraint equations yield
to the following expression for the quintessence field:
\begin{eqnarray}
\label{Q
sol}Q(t)=Q_0\left(\frac{t}{t_0}\right)^{\frac{1+(N-1)\beta}{1-\beta}}
\end{eqnarray}
where $Q_0=\sqrt{\frac{6k(1+w_{\sst{Q}})(\beta-2)(1-\beta)^2\Phi_0
{t_0}^2} {8\pi(2\beta-3(1+w_{\sst{Q}}))(1+(N-1)\beta)^2{a_0}^2}}$.
From expressions (\ref{rhoquinta}) and (\ref{pquinta}), together
with the equation of state for the quitessence field yields to
\begin{eqnarray}
\label{VQ
sol}V(Q)=V(Q_0)\left(\frac{Q}{Q_0}\right)^\frac{2N\beta}{1+(N-1)\beta}
\end{eqnarray}
where $V(Q_0)=\frac{6k(\beta-2)(1-w_{\sst{Q}})\Phi_0}
{16\pi(2\beta-3(1+w_{\sst{Q}})){a_0}^2}$. Notice that, since
$\beta <0$, then we need just only to satisfy $N>2$ in order to
get an usual quintessence scalar potential, i.e., an inverse power
law scalar potential.

\section{Transport equation for bulk viscosity}

At this point we word like to describe dissipation due to bulk
viscosity $\xi$, via Israel-Stewart theory. The bulk viscous
pressure $\Pi$ is given by the transport equation (linear in
$\Pi$)
\begin{eqnarray}
\label{8} \tau \dot{\Pi}+\Pi =-3 H \xi  -\frac{1}{2} \tau \Pi
\left(3H+\frac{\dot{\tau}}{\tau}-\frac{\dot{\xi}}{\xi}-\frac{\dot{T}}{T}\right),
\end{eqnarray}
where $\tau$ is the relaxation time (which removes the problem of
infinite propagation speeds) and $T$ is the temperature of the
fluid. In the non-causal formulation $\tau=0$ and then
Eq.(\ref{8}) has a simple form $\Pi=-3 H \xi$.

Following refs.\cite{Belinskii,4} we will take the  thermodynamic
quantities to be simple power functions of the matter density
$\rho_{_{\sst{M}}}$,
\begin{eqnarray}
\label{suposiciones} \xi = \alpha {\rho_{\sst M}}^{m} \;\;\;, T=
\mu {\rho_{\sst M}}^r \;\;\; y \;\;\; \tau
=\frac{\xi}{\rho_{_{\sst{M}}}}= \alpha {\rho_{\sst M}}^{m-1}
\end{eqnarray}
with $\alpha$,$\mu$,$m$ and $r$ greater them zero. The expression
for $\tau$ is used as a simple procedure to ensure that the speed
of viscous pulses does not exceed the speed of light. For an
expanding cosmological model the constant $m$ should be positive
and we should satisfy
\begin{eqnarray}
\label{condicion para tau} \tau > H^{-1},
\end{eqnarray}
in order to have a proper physical behavior  for $\xi$ and $\tau$
\cite{4,Banerjee}.


>From the expansion law(\ref{factor de escala}) and from
eqs.(\ref{rhofluido}), (\ref{8}) and(\ref{suposiciones}) (with the
standard  relation for the temperature of a barotropic fluid in
which  $r=(\gamma-1)/\gamma$) we obtain
\begin{eqnarray} \label{27}
\ddot{\rho}_{\sst M}+(3N+1)\frac{\dot{\rho}_{\sst
M}}{t}+\frac{1}{\alpha}{\rho}^{1-m}_{\sst M}\dot{\rho}_{\sst
M}+\frac{3N \gamma}{\alpha t}{\rho}^{2-m}_{\sst M} \nonumber
\\ -9N^2\left(1-\frac{\gamma}{2}\right)\frac{\rho_{\sst
M}}{t^2}-\frac{(2\gamma-1)}{2\gamma}\frac{\dot{\rho}^{2}_{\sst
M}}{\rho_{\sst M}}=0.
\end{eqnarray}

This differential eq. may be solve for some specific values of the
$m$ parameter. Taking the following anzats\cite{Banerjee}
\begin{eqnarray}
\label{anzats} \rho_{_{\sst{M}}}=\rho^{0}_{_{\sst{M}}}
\left(\frac{t}{t_0}\right)^{(\frac{-1}{1-m})},
\end{eqnarray}
with $m<1$ in order that $\rho_{\sst{M}}$ decreases when $t$
increase, we obtain from eq.(\ref{27}) the constant
$\rho_{\sst{M}}^{o}$ becomes given by
\begin{eqnarray}
\rho^0_{_M}=\left( \frac{\alpha}{t_0 (1-m)} \right) \times
\,\,\,\,\,\,\,\,\,\,\,\,\,\,\,\,\,\,\,\,\,\,\,\,\,\,\,\,\,\,\,\,
{}^{}
\\ \nonumber \left(
\frac{(2\gamma)^{-1}-3N(1-m)-9N^2(1-\gamma/2)(1-m)^2}{(1-3N\gamma(1-m
))}\right)^{1/(1-m)}.
\end{eqnarray}
From this we can find $\alpha$ as a function of the parameters
$\gamma$, $N$, $m$ and $\rho^0_{_{\sst{M}}}$. The pressure
$p_{\sst{M}}$ is obtained using the equation of state $p_{\sst{M}}
= (\gamma - 1) \rho_{\sst{M}}$.

Now, from eq.(\ref{rhofluido}) and the barotropic eq. of state for
the matter fluid we obtain for the bulk viscous pressure
\begin{eqnarray}\label{shear viscosidad}
\Pi= -\frac{\dot{\rho}_{_{\sst{M}}}}{3H}- \gamma
\rho_{_{\sst{M}}},
\end{eqnarray}
and substituting eq.(\ref{anzats}) in this latter eq. we get
\begin{eqnarray}
\label{final de viscosidad} \Pi(t)=
-\left(\gamma-\frac{1}{3N(1-m)} \right) \rho^0_{_{\sst{M}}}
\left(\frac{t}{t_0}\right)^{-1/(1-m)}.
\end{eqnarray}
We see that $\Pi<0$ if $\gamma>1/(3N (1-m))$, $\Pi > 0$ if
$\gamma<1/(3N (1-m))$ and $\Pi=0$ if $m=1-1/(3N\gamma)$.

In order to satisfy the field eqs. of motion the parameters should
follow the equality $\frac{1}{1-m}=2-\frac{2N}{1-\beta}$, with
$\beta$ expressed in terms of $w_{\sst{Q}}$, thus $m$ becomes a
function of the parameter $N$ and $w_{\sst{Q}}$,
\begin{equation}
\ds m = 1-\frac{1}{2}\frac{1-\beta}{1-\beta - N} \label{m}
\end{equation}
where $\beta$ is a function of the $w_{\sst{Q}}$ parameter.

At local equilibrium, the entropy  satisfy the Gibbs equation
\begin{eqnarray} \label{32}
TdS=(\rho_{\sst{M}}+p_{\sst{M}})d\left(\frac{1}{n}\right)+\frac{1}{n}
d\rho_{\sst{M}},
\end{eqnarray}
where $n$ is the number density and satisfy the conservation
equation
\begin{eqnarray} \dot{n}+3Hn=0,
\end{eqnarray}
from which we get
\begin{eqnarray} \label{34}
n(t)=n_{0}\left(\frac{t}{t_{0}}\right)^{-3N}
\end{eqnarray}
or equivalently $n(t) a^{3}(t)= const$.

Using(\ref{32}) and(\ref{34}) we obtain the well known evolution
equation for the entropy (neglecting the heat flux and the shear
viscosity)
\begin{eqnarray} \label{35}
\dot{S}=-\frac{3H\Pi}{nT},
\end{eqnarray}
 thus from eqs.~
(\ref{pquinta}), (\ref{suposiciones}), (\ref{anzats}), (\ref{final
de viscosidad}) and(\ref{34})
we get
\begin{eqnarray} \label{Spunto}
\dot{S}=\frac{3n\gamma-1/(1-m)}{n_{_{0}}\mu t_{_{0}}}
(\rho^0_{_{\sst{M}}})^{1/\gamma}
\left(\frac{t}{t_0}\right)^{(3N-1)-1/\gamma(1-m)}.
\end{eqnarray}
For proper physical behaviour we must satisfy $\dot{S}>0$, which
implies the condition $m<1-1/\gamma(3N-1)$. This inequality
together with eq.(\ref{m}) give an inequality for $\gamma$ as
function of the parameters $N$ and $w_{\sst{Q}}$, $$ \ds \gamma >
2 \frac{(1-\beta -N)}{(1-\beta)(3 N -1)}. $$ Thus, this parameter
becomes bounded from below.

The total entropy $\Sigma$ in a comoving volume is defined by
means $\Sigma=S n a^{^{3}}(t)$. Then, by taking  Eq.(\ref{34})
and(\ref{35}) we may write the growth of total nondimensional
comoving entropy over a proper time interval $t_f$ and $t_i$
as\cite{4}
\begin{eqnarray}
\label{total comoving entropy}
\Sigma_f-\Sigma_i=-\frac{3}{k_{_{B}}} \int^{t_f}_{t_i} \frac{\Pi H
a^{^{3}}(t)}{T} dt.
\end{eqnarray}

Now taking  $t_i$ and $t_f$ as the beginning and the exit time for
the inflation period of the universe respectively and from
eqs.(\ref{factor de escala}), (\ref{suposiciones}),
(\ref{anzats}), (\ref{final de viscosidad}), and (\ref{total
comoving entropy}) we obtain for the increase the total
nondimensional entropy in the comoving volume $a^{3}(t)$ the
following expression
$$ \hspace{-3.50cm}\label{Dedo} \Sigma_f-\Sigma_i=\frac{\gamma
a^3_0 \, (\rho^0_{{_M}})^{1/\gamma}}{k_{_{B}} \mu} \times
$$
\begin{eqnarray}\label{Dedo}
\hspace{1.50cm}\left[\left(\frac{t_f}{t_0}
\right)^{3N-\frac{1}{\gamma(1-m)}} - \left(\frac{t_i}{t_0}
\right)^{3N-\frac{1}{\gamma(1-m)}} \right].
\end{eqnarray}
We take the values for the beginning and ending time of inflation
as $t_i \approx 10^{-35}$ s and $t_f \approx 10^{-32}$ s,
respectively. In the following numerical calculation we  take the
reference time $t_0$ equal to  $t_f$. Considering that the
universe at the end of the period inflation  exits to the
radiation era, we can constraint some of the constants of
integration of the above formulae. Effectively, it is known that
the temperature of the universe at the beginning of the radiation
era is approximately $T \sim 10^{14} \, GeV \approx 1.16 \times
10^{27}$ K. Thus, the temperature at the end of inflation must be
$T_f=1.16 \times 10^{27}$ K. On the other hand, we know that in
this period $\rho=a_r\, T^4$, where $a_r=\frac{\pi^2k^4_{_B}}{15
c^3 \hbar^3}\approx 7.56 \times 10^{-15} \, J \, m^{-3} K^{-4}$.
From here we conclude that at the end of inflation any period (or
at the beginning of the radiation era) we have $\rho_{_{f}}
\approx 10^{93} \, J \, m^{-3}$. This implies that  $r=1/4$ for
the exponent in the temperature $T$, i.e. $\gamma=4/3$, as we can
see from eq.~\ref{suposiciones}.

Thus the typical values for the inflationary period are\cite{4}:
\begin{eqnarray}
\label{values} t_i \approx 10^{-35} \, s; \,\,\, t_f \approx
10^{-32} \, s; \,\,\, 
a_i \approx c t_i, \nonumber \\ T_f \approx 10^{27} \, K, \,\,\,
\gamma=4/3, \,\,\, \rho \approx \times 10^{93} J/m^3.
\end{eqnarray}
The e-folding parameter $Z=ln[a(t_f)/a(t_i)]$ for the power law
inflation(\ref{factor de escala}) takes the form
\begin{eqnarray}
\label{Z para power law} Z=n \, ln\left(\frac{t_f}{t_i} \right).
\end{eqnarray}
It is well known that for solving most the problems of the
standard model in cosmology we must have $Z \approx 60-70$.
Thus from(\ref{suposiciones}), (\ref{Dedo}), (\ref{values})
and(\ref{Z para power law}) we have
\begin{eqnarray}
\label{Sigma para tn} \Sigma_f-\Sigma_i \approx \frac{4 a_i \,
e^{3Z}
(\rho_{_{\sst{M}}}^f)^{3/4}}{3 k_{_{B}} \mu} 
\left(1- (10^{-3})^{3N-\frac{3}{4(1-m)}} \right),
\end{eqnarray}
where we have used the relation $a^3_f=a_i^3 e^{3Z}$, following
from the definition of $Z$, eq.\ref{Z para power law}. We see from
Eq.(\ref{Sigma para tn}) that if the inequality $3N \gamma
(1-m)>>1$ is satisfied (in this case $\Sigma_f >> \Sigma_i$) we
obtain the accepted value for the total entropy in the observable
universe\cite{Ko}
\begin{eqnarray} \label{valor sigma}
\Sigma \approx \times 10^{88}.
\end{eqnarray}
Then this model can account for the generally accepted entropy
production via causal dissipative inflation, \emph{without
reheating}.

Finally, for complete, we will see what happen when
 $m=1$. For this value and considering
eq.(\ref{27}),  which gives the solution\cite{Banerjee}
\begin{eqnarray}
\rho_{_{\sst{M}}}=\rho^0_{_{\sst{M}}}  e^{-2\gamma t/\alpha}
\left( \frac{t}{t_0}\right)^{-\gamma(3N+2)},
\end{eqnarray}
where $N^2=2\gamma/9$ and $\rho^0_{_{\sst{M}}}$ is an arbitrary
constant. Unfortunately this solution implies a decreasing entropy
\begin{eqnarray}
S(t)=\frac{\gamma}{\mu_0 n_0} \left(\rho^{^{0}}_{_{\sst{M}}}
\right)^{1/\gamma} \left( \frac{t_0}{t} \right)^2 e^{-2 t/\alpha}+
const
\end{eqnarray}
which certainly we should neglect in order to finish with a
reasonable physical model.

\section{Conclusions}

To conclude, on the light of the above results we may say that
there exist an intimate relation between causal dissipative
inflation and reheating. We expect to come back to this point of
research in the near future, where extensions to more general
universe models will be applied.

\section{\bf Acknowledgements}
This work is dedicated to Alberto Garc\'{\i}a's 60$^{th}$
birthday. SdC and MC were supported by COMISION NACIONAL DE
CIENCIAS Y TECNOLOGIA through Grants FONDECYT N$^0$ 1030469 and
N$^0$ 1010485, respectively. Also, SdC  was supported by PUCV-DI
grant N$^0$ 123.764/03, MC by Direcci\'{o}n de Promoci\'{o}n y
Desarrollo de la Universidad del B\'{\i}o-B\'{\i}o and FP was
supported from DIUFRO N$^0$ 20228.


\end{document}